\documentclass[%
 aip,
 amsmath,amssymb,
 reprint,%
]{revtex4-1}
\usepackage{amsmath}
\usepackage{dcolumn}
\usepackage{epsfig}
\usepackage{epstopdf}
\usepackage{graphicx}
\usepackage{amsfonts}
\usepackage{mathrsfs}
\usepackage{color}
\usepackage{multirow}
\usepackage{hyperref,natbib}
\usepackage{graphicx}
\usepackage{dcolumn}
\usepackage{bm}
\usepackage[utf8]{inputenc}
\usepackage[T1]{fontenc}
\usepackage{etoolbox}
\usepackage[letterpaper,margin=1in]{geometry}
\makeatletter
\def\@email#1#2{%
 \endgroup
 \patchcmd{\titleblock@produce}
  {\frontmatter@RRAPformat}
  {\frontmatter@RRAPformat{\produce@RRAP{*#1\href{mailto:#2}{#2}}}\frontmatter@RRAPformat}
  {}{}
}%
\makeatother

\preprint{AIP/123-QED}

\begin{document}

\title{Single-atom sensor for low-frequency electric field}
\newcommand{\APM}{Wuhan Institute of Physics and Mathematics, Innovation Academy of Precision Measurement Science and Technology, Chinese Academy of Sciences, Wuhan 430071, China}
\newcommand{\UCAS}{University of Chinese Academy of Sciences, Beijing 100049, China}
\newcommand{\GZII}{Guangzhou Institute of Industrial Intelligence, Guangzhou 511458, China}
\newcommand{\GZIIT}{Research Center for Quantum Precision Measurement, Guangzhou Institute of Industry Co., LTD, Guangzhou, 511458, China}
\newcommand{\ZJNU}{Department of Physics, Zhejiang Normal University, Jinhua 321004, China}
\newcommand{\HNU}{Key Laboratory of Low-Dimensional Quantum Structures and Quantum Control of Ministry of Education,
Department of Physics and Synergetic Innovation Center for Quantum Effects and Applications, Hunan Normal University, Changsha 410081, China}
\author{Quan Yuan}
\thanks{Co-first authors with equal contribution}
\affiliation{\GZII}
\affiliation{\APM}

\author{Shuang-Qing Dai}
\thanks{Co-first authors with equal contribution}
\affiliation{\APM}
\affiliation{\UCAS}

\author{Tai-Hao Cui}
\affiliation{\APM}
\affiliation{\UCAS}
\author{Pei-Dong Li}
\affiliation{\APM}
\affiliation{\UCAS}
\author{Yuan-Zhang Dong}
\affiliation{\APM}
\affiliation{\UCAS}
\author{Zhuo-Zhu Wu}
\affiliation{\APM}
\affiliation{\UCAS}
\author{Ji Li}
\affiliation{\GZIIT}
\author{Fei Zhou}
\affiliation{\APM}
\affiliation{\GZIIT}
\author{Jian-Qi Zhang}
\email{changjianqi@gmail.com}
\affiliation{\APM}
\author{Liang Chen}
\email{liangchen@wipm.ac.cn}
\affiliation{\APM}
\affiliation{\GZIIT}
\author{Mang Feng}
\email{mangfeng@wipm.ac.cn}
\affiliation{\APM}
\affiliation{\GZIIT}
\affiliation{\HNU}
\affiliation{\ZJNU}


\begin{abstract}
Precision measurement of low-frequency electric field (LFEF) signals with frequency from 30 kHz to 300 kHz is crucial for advancing both fundamental science and practical applications, owing to their unique frequency regime. For conventional electromagnetic antennas, the long wavelength (i.e., several kilometers) of the LFEF leads to a severe size constraint that efficient radiation becomes challenging to achieve when the antenna size is much smaller than the long wavelength of the LFEF signals, which in turn results in a reduction of measurement sensitivity and compromises antenna's performance. By exploiting the high intrinsic sensitivity of cold trapped ions to weak alternating electric signals via Coulomb interaction, we demonstrate a single-ion phonon laser sensor acted by an injection-locked $^{40}$Ca$^{+}$ ion confined in a surface-electrode trap. Combining the beat frequency technique with the injection-locked phonon laser oscillation, we demonstrate a practical and efficient approach for simultaneous extraction of the frequency, phase, and amplitude from a single measurement, without the need for sideband cooling. This approach achieves precision detection for LFEF signals with the sensitivity of 403.8 $\mu$V/($m\cdot{\mathrm{Hz}^{1/2}}$) and the detection limit of 61.5 $\mu$V/m. Besides, this approach also shows remarkable robustness against noise. Our study helps realizing practical single-atom sensors in the low-frequency regime, opening avenues for applications in subsurface communication, precision metrology, mass spectrometry, and biomedical monitoring.
\end{abstract}
\maketitle

\vspace{0.5cm}




The precision measurement of weak alternating signals is an important topic in physics and metrology. Such signals often carry critical information about new physical phenomena, ranging from ultralight dark-matter fields~\cite{tretiak2022improved} and geophysical transients~\cite{li2013review} to neuronal oscillations~\cite{buzsaki2004neuronal}. Their detection is hindered by weak amplitudes, low frequencies, and environmental noise. In particular, precision detection of low-frequency electric fields (LFEFs) between 30 kHz and 300 kHz is of great interest because their long wavelength and low propagation loss  are vital for subsurface and underwater wireless communications \cite{9905692,6197530}. For conventional antennas, effective coupling to the long wavelength (i.e, several kilometers) of a LFEF requires the antenna size comparable to the signal wavelength, and severe size-wavelength mismatch impairs signal detection capability. The inherent constraint of conventional antennas for LFEFs severely compromises the detection sensitivity of miniaturized antennas, fundamentally hindering the miniaturization of practical LFEF detection systems.

To overcome this difficulty, various measurement techniques have been developed to achieve weak signal detection~\cite{poor2013introduction,huang2023review}, one of which is the beat frequency technique. This technique converts two nearby frequencies into a detectable low-frequency signal, providing a powerful approach for processing optical and electrical signals~\cite{diebold2013digitally}. It has been extensively applied in optical~\cite{doi:10.1126/science.150.3693.149,doi:10.1126/science.163.3865.345,RevModPhys.87.637,Zemanek:94,Greiner:98,10.1038/nphoton.2014.85}, atomic~\cite{1446564,jing2020atomic,photonics9040250,PhysRevApplied.19.034078}, and electronic systems~\cite{8551288,7000572,wang2017micro,Kulkarni2014} for communication demodulation \cite{10.1038/s41567-020-0918-5,Lucy:67,6007062}, biomedical signal detection \cite{SALMEEN19721172,Zhang:11,10022313,9174987}, and precision measurement \cite{Dai:12,GAO2013137,10.1038/nphoton.2013.245}.

In particular, applying the beat frequency technique in laser systems has given rise to various advanced detection techniques, including laser heterodyne detection \cite{protopopov2009laser,Jin2021,Liang2022,doi:10.1126/sciadv.adp8556}, dual-comb spectroscopy \cite{Coddington:16,10.1038/nphoton.2009.217,10.1038/nphoton.2015.250,doi:10.1126/science.ads6292}, and fiber Bragg grating \cite{10.1063/1.1148392,618320,9174987,GUO2023103155}.  These techniques represent powerful applications of the beat frequency principle in the optical domain, but not directly applicable to the LFEF detection. Extending the exquisite sensitivity of optical beat frequency techniques to the direct detection of LFEFs remains challenging for conventional photonic systems, primarily due to their susceptibility to environmental noise and limited electro-optic response at these frequencies.

Here, we present a fundamentally distinct LFEF measurement technique for precision detection of electric fields using a trapped ion. As ions are charged and interact with external electric fields via the Coulomb force, they exhibit an high sensitivity to  fluctuation of external electric fields due to their extremely high charge-to-mass ratio ~\cite{RevModPhys.87.1419,RevModPhys.89.035002,doi:10.1126/science.abb2823,PhysRevApplied.19.064062}. This intrinsic property allows ion-based electric field sensors to break through the conventional size constraints and be miniaturized down to the micrometer scale with high sensitivity detection for LFEF signals. In our experiment, the trapped ion functions as an injection-locked single-ion phonon laser, the vibrational analog of an optical laser, exhibiting amplification \cite{PhononLaser} and frequency stabilization via an injection-locking signal \cite{InjectionLocking}.

By applying the beat frequency technique to the input LFEF signal and the injection-locked phonon laser, we coherently map the input signal onto the ion's vibration, enabling simultaneous extraction of its frequency, phase, and amplitude in a single measurement. This approach provides high sensitivity and resolution in the kHz regime, while requiring neither sideband cooling nor special operating conditions. It thus offers a compact and practical route to highly sensitive weak-signal sensing.
\\

\begin{figure}[t]
\centering
\includegraphics[width=0.48 \textwidth]{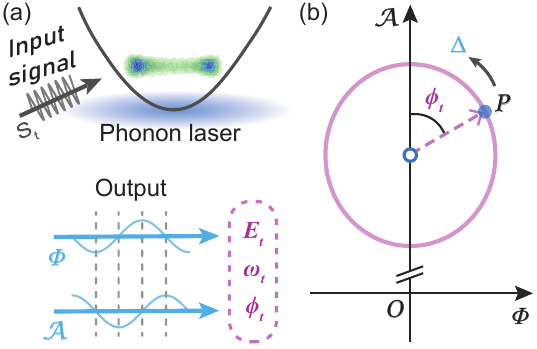}
\caption{Principle of the beat frequency measurement by single-ion phonon laser.
	(a) The scheme of the beat frequency measurement by single-ion phonon laser, where the vibrational motion of the single ion is modulated by the input signal $S_t$, and the responses are characterized by the evolution of the phase $\mathit{\Phi}$ and amplitude $\mathcal{A}$ of the phonon laser. (b) Trajectories in phase-amplitude coordinate of the phonon laser. }
\label{fig:1}
\end{figure}

\noindent
Our experiment is carried out in a surface-electrode trap (SET) with the measured secular frequency along z-axis of $\omega_{z}/2\pi$ = 238.42(0.01) kHz. A single $^{40}$Ca$^+$ ion, located at 800 $\mu$m above the center of the SET, behaves as a phonon laser in the direction of $\omega_{z}$ under the irradiation of two detuned 397-nm laser beams, where one is red-detuned with $\Delta_{r}$ = -115 MHz and the other is blue-detuned with $\Delta_{b}$ = 30 MHz. We saturate the laser power with the ratio $r = s_{b}/s_{r}$ = 0.58, where $s_{r(b)}$ represents the saturation parameter of the red (blue)-detuned laser beam. Throughout the whole experiment, an additional over-saturated 866-nm laser beam is always on for repumping. This setup ensures the generation of a stable phonon laser system with the oscillation amplitude of 25.5 $\mu$m. In particular, only Doppler cooling of the trapped ions is required in our scheme.

For our purpose, we employ the arbitrary waveform generator (AWG) to apply the injection locking signal $S_i=V_i\sin(2\pi\omega_it)$ to the axis electrode of the SET, and inject the synchronization signal into the time-to-amplitude converter \cite{SM}. The injection intensity is fixed to be $V_i$ = 7.5 mV with the frequency set to be the same as $\omega_z$. As a result, the frequency of the single-ion phonon laser is locked and synchronized with $S_i$ accompanied with an amplified oscillation. The LFEF signal to be detected is also generated from the AWG in the form of $S_t=V_t\sin(2\pi\omega_tt+\phi_t)$ and applied to the axis electrode (see Fig. \ref{fig:1}(a)). Thus, the LFEF signal applied to the trapped ion can be denoted as $E_t=kV_t$ with a response parameter $k$, which is mainly determined by the materials and geometric structure of the axis electrode. We denote the input signal as $(V_t,\Delta,\phi_t)$, where $\Delta=\omega_t-\omega_i$ is the beat frequency between $S_i$ and $S_t$.

To perform the beat frequency measurement of LFEF signals, we combine two waves of slightly different frequencies, which gives rise to beats at their frequency difference. In our experiment, we utilize the external vibrational motion of the injection-locked phonon laser system to perform beat frequency measurements of LFEF signals. The beat frequency between the injection-locked phonon laser and the input signal $S_t$ generates a modulation of the phonon laser's amplitude and phase. By synchronously monitoring the phonon laser's vibration through photon scattering, we extract the frequency, phase, and amplitude of the input signal $S_t$ from the beat-induced evolution of the phonon laser.

\begin{figure}[t]
\centering
\includegraphics[width=0.4\textwidth]{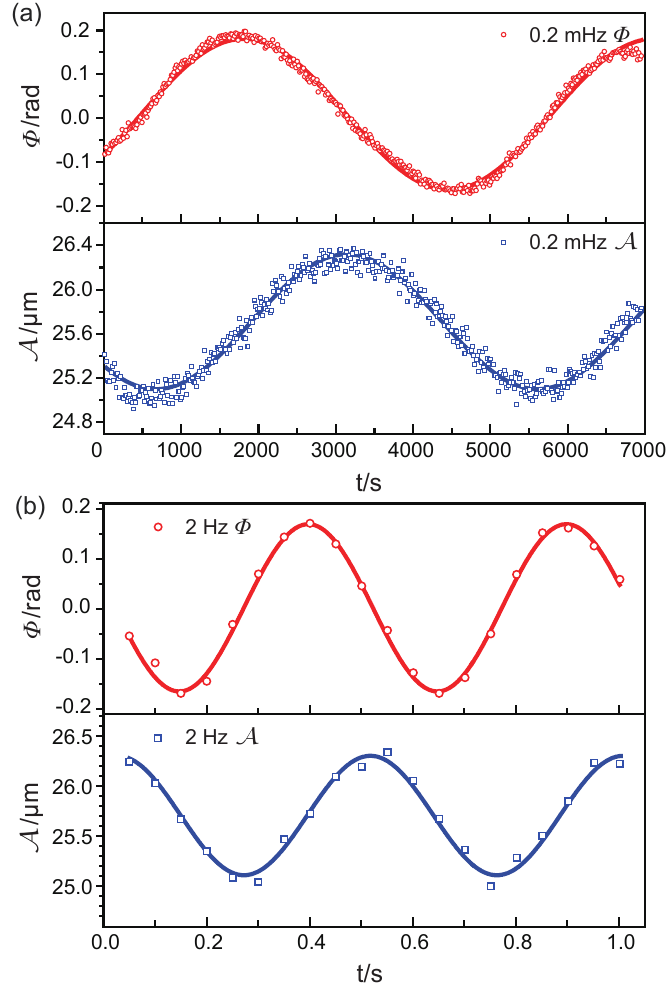}
\caption{Beat frequency measurement of sinusoidal input signals. Evolution of the measured phase $\mathit{\Phi}$ (red) and amplitude $\mathcal{A}$ (blue) of the phonon laser with input signal of (a) 0.2 mHz and (b) 2 Hz. The red and blue lines are the sinusoidal fitting results of the data points.}
\label{fig:2}
\end{figure}

To show the principle in more details, we describe the oscillation displacement of the single-ion phonon laser as $Z=\mathcal{A}\sin(2\pi\omega_it+\mathit{\Phi})$, with the slowly varying amplitude $\mathcal{A}$ and phase $\mathit{\Phi}$. We have the motion equation of the ion interacting with two laser beams along the z-axis of the trap, with both $S_i$ and $S_t$ applied,
\begin{equation}
	m\ddot{Z}=-m\omega_z^2Z+F_c+F_g+F(t) \label{1},
\end{equation}
where $m$ is the mass of the ion, $Z$ is the displacement of the ion, $F_c$ ($F_g$) represents the scattering force of the cooling (gain) laser beam, and $F(t)=ke(S_i+S_t)=ke[V_i\sin(2\pi\omega_it)+V_t\sin(2\pi\omega_tt+\phi_t)]$ corresponds to the force of the applied signals acting on the ion with the response parameter $k$. A beat frequency emerges from the superposition of the phonon laser and $S_t$, manifesting as periodic fluctuations on both amplitude and phase of the phonon laser.
    Here we assume $V_t\ll V_i$ and $|\Delta|\ll\omega_i$ and the outputs of $\mathit{\Phi}$ and $\mathcal{A}$ vary periodically with $\Delta$ as \cite{SM}
\begin{align}
		\mathit{\Phi}&\approx-R_pkV_t\sin(\Delta t+\phi_t), \label{2}\\
		\mathcal{A}&\approx A_0+R_AkV_t\cos(\Delta t+\phi_t), \label{3}
\end{align}
where the response factors $R_p=e/2m\omega_i\Delta A_0$ and $R_A=e/2m\omega_i\Delta$, with $A_0$ the average amplitude of the injection-locked phonon laser without the input signal applied.

To illustrate the response of the phonon laser system to the input signal more vividly, we decompose Eqs. (\ref{2}) and (\ref{3}) in the phase-amplitude coordinate, as illustrated in Fig. \ref{fig:1}(b). Initially the phonon laser lies at the position of $(0,A_0)$, corresponding to the locked phase and the amplified amplitude under the injection locking. After the input signal is applied, the phonon laser is dragged to P along a direction with an angle of $\phi_t$ to the positive $\mathcal{A}$-axis, and then begins to rotate around $(0,A_0)$ at a frequency of $|\Delta|$ along the elliptical path. A positive (negative) $\Delta$ leads to a counterclockwise (clockwise) rotation. 

Experimentally, we record the distribution of accumulated photon counts as a function of their arrival time counted by a photomultiplier tube (PMT) and a time-to-amplitude converter. The phase $\mathit{\Phi}$ and amplitude $\mathcal{A}$ of the phonon laser can be acquired by fitting the accumulated photon counts with the convolution of the scattering rate of the phonon laser \cite{Liu2021PhononLaser,Wei2022DCelectric}. The accumulation time can be set, according to $\Delta$, to meet the sampling theorem.  By analyzing the continuously accumulated curves regarding the photon scattering rate, $\mathit{\Phi}$ and $\mathcal{A}$ in different moments can be obtained. Then the variations of $\mathit{\Phi}$ and $\mathcal{A}$, evolving synchronically with $S_t$, can be fitted by Eqs. (\ref{2}) and (\ref{3}) and thus, we acquire the characteristics of $S_t$. In our scheme, a higher sampling rate for sinusoidal fitting results in a shorter accumulation time per curve.

Fig. \ref{fig:2} presents the experimentally obtained $\mathit{\Phi}$ and $\mathcal{A}$ for (a) $\Delta$ =  0.2 mHz and (b) $\Delta$ = 2 Hz. These results clearly demonstrate the beat frequency between the phonon laser and the input signal. Moreover, it is evident that both the uncertainties in the measurements and the errors of $\mathcal{A}$ in the fitting curve are significantly higher than those of $\mathit{\Phi}$. This results from the phase locking and limitation of the fluorescence acquisitions. Therefore, in the following experiments, we would fit the measured $\mathit{\Phi}$ to Eq. (\ref{2}) with a sinusoidal function using $\mathit{\Phi}_{fit}=-E_{\mathrm{fit}}\sin(\Delta_{\mathrm{fit}}t-\phi_{\mathrm{fit}})$ to extract the fitting result $E_{\mathrm{fit}}, \Delta_{\mathrm{fit}}$, and $\phi_{\mathrm{fit}}$, where the parameter of the input signal can be derived. However, since the evolution of $\mathit{\Phi}$ is the same for both cases of $(V_t,-\Delta, \pi - \phi_t)$ and  $(V_t,\Delta,\phi_t)$, we cannot identify the input signal from $\mathit{\Phi}$ alone. As such, we use the amplitude $\mathcal{A}$ to provide the necessary discrimination.

By precalibrating how the input signal affects the evolution range of $\mathit{\Phi}$ (which will be clarified later in the sensitivity of the amplitude measurement), we can directly convert the fitted amplitude $\phi_{fit}$ into $\phi_t$ due to their one-to-one mapping. For example, we fit Eq. (\ref{2}) with sinusoidal functions to the data in Fig. \ref{fig:2}(a,b) and extract the input signal parameters: (a)  (1.26 mV, 0.2 mHz, 0.82$\pi)$ and (b) (1.20 mV, 2 Hz, -0.11$\pi$).

We next discuss the sensitivity and limitations of our protocol for simultaneously detecting the three characteristics of the input signal. Note that the measured frequency difference $\Delta_{fit}$ directly reflects the properties of $S_t$, while the amplitude and phase of the input signal are converted from the measured radian amplitude $E_{fit}$ and phase $\phi_{fit}$.
We define the sensitivities of frequency, phase and amplitude by the minimum detectable change $\sigma_D$, $\sigma_p$, and $\sigma_a$, i.e., the measurement uncertainties of $\Delta_{fit},E_{fit},\phi_{fit}$, within the detection bandwidth $1/\sqrt{t_{tot}}$, as evaluated later from experimental results. Thus they are given by $\eta_D=\sigma_{D}\sqrt{t_{tot}}$, $\eta_p=\sigma_{p}\sqrt{t_{tot}}$ and $\eta_a=\sigma_{a}\sqrt{t_{tot}}/(\partial E_{fit}/\partial E_t)$, respectively, where  $t_{tot}$ is the total measurement time.\cite{Liu2021PhononLaser,Wei2022DCelectric}
	
\begin{table}[b]
	\begin{center}
		\caption{Comparison of the beat frequency measurements as well as the fitting error (FE) from sinusoidal fitting, where the input signal is set as (1.25 mV, $\Delta$, 0) and total measurement time is $t_{tot}\ge 1/\Delta$ for each data point. For example, we set $t_{tot}$ = 2100 s for $\Delta$ = 0.5 mHz, and $t_{tot}$ = 2 s for $\Delta$ = 1 Hz. }
		\begin{tabular}{c|c|c}
			\hline
			\hspace{1.5em}$\Delta$~(Hz)\hspace{1.5em} &\hspace{1.5em}$\Delta_{fit}$~(Hz)\hspace{1.5em}
			&\hspace{1.5em}FE~(Hz)\hspace{1.5em} \\
			\hline
			2e-4	&1.8845e-4	 	&-3.22e-7 \\
			3e-4	&3.1651e-4	 	&-1.18e-6 \\
			5e-4	&5.0281e-4	 	&-2.19e-6 \\
			1e-3	&1.0004e-3	    &-4.54e-6 \\
			2e-3	&1.9963e-3	  	&-3.19e-6 \\
			5e-3	&5.0273e-3	   	&-1.61e-5 \\
			0.01	&0.0101	 	    &-4.16e-5 \\
			0.02	&0.0204	 	    &-1.21e-4 \\
			0.05	&0.0514	 	    &-3.87e-4 \\
			0.1	 	&0.1020	 	    &-6.06e-4 \\
			0.2	 	&0.2033	 	    &-2.15e-3 \\
			0.5	 	&0.5177	 	    &-4.48e-3 \\
			1	 	&1.0073	        &-5.13e-3 \\
			2	 	&1.9996	 	    &-1.16e-2 \\
			\hline
		\end{tabular}
	\end{center}\label{table:1}
\end{table}

In Table I, we present the results of (1.25 mV, $\Delta$, 0) with $\Delta$ varying for a quite wide range from 0.2 mHz to 2 Hz. Obviously, each data point provides a good reconstruction of the input signal with $\Delta_{fit} \approx \Delta$, unequivocally demonstrating the overall reliability of our scheme in this frequency range. With the reduction of $\Delta$, the fitting error (FE), denoting the uncertainty obtained by fitting the measured phase of the phonon laser to sinusoidal function, decreases because more samples are collected over a longer period, while the accuracy also slightly decreases due to the long-term drift of $\omega_z$ \cite{SM}.

By elaborately regulating the electrode voltages of the SET, we can vary $\Delta$ to shorten the required total measurement time. Theoretically the phonon laser system can respond to a quite wide range of frequency as long as $|\Delta|\ll\omega_{z}$ is satisfied. Experimentally, however, our frequency resolution is limited by the sampling rate and system stability. A larger $\Delta$ requires less accumulating time, which is unfavorable to the accuracy of the fitting. As for the other hand, a smaller $\Delta$ leads to a longer total measurement time for ensuring the fitting accuracy, which requires highly long-term stability of the system. At a frequency near the middle of the range (e.g., $\Delta$ = 0.05 Hz) and for $t_{tot}$ = 20 s, $\sigma_D$ = 0.75 mHz, the optimal frequency sensitivity can be obtained at 3.4 mHz/$\sqrt{\mathrm{Hz}}$.

\begin{figure}[t]
\centering
\includegraphics[width=0.48\textwidth]{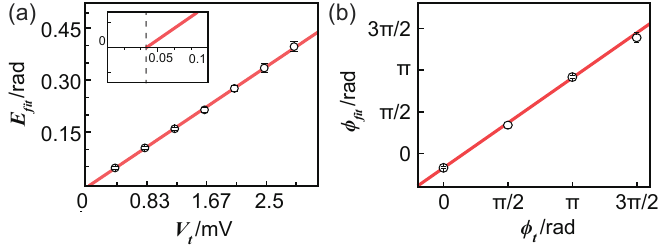}
\caption{ Evaluation of the accuracy of $\mathit{\Phi}$. (a) The fitting amplitude $E_{fit}$ of $\mathit{\Phi}$ versus the given amplitude $V_t$ of the input signal ($V_t$, 0.05 Hz,0), with $V_t$ varying from 0.42 mV to 2.92 mV. The red line is a linear fit of the data points with the slope of $\partial E_{fit}/\partial V_t$ = 136.8 mrad/mV. (b) The fitting phase $\phi_{fit}$ of $\mathit{\Phi}$ versus the given phase $\phi_t$ of the input signal (1.25 mV, 0.05 Hz, $\phi_t)$. Error bars are statistical standard errors from 10 measurements for each data point with $t_{tot}$ = 20 s for each measurement. } \label{fig:3}
\end{figure}

The fitting amplitude $E_{fit}$ varies linearly with the given $V_t$, as plotted in Fig. \ref{fig:3}(a). In our experiment, the response parameter $k$ is calculated as 1.3 $m^{-1}$ with the fitted slope of $\partial E_{fit}/\partial V_t$ = 136.8 mrad/mV and $\omega_{i}/2\pi$ = $\omega_{z}/2\pi$ = 238.42 kHz. Due to the inherent noises, the fitted line does not pass through the origin but approaches zero at $V_t$ = 46.2 $\mu$V, indicating the minimum detectable strength of the LFEF in our system is 61.5 $\mu$V/m. Unlike the one-to-one reconstruction in frequency as shown above, the amplitude sensitivity can be improved by increasing the slope $\partial E_{fit}/\partial V_t$. According to Eq. (\ref{2}), the slope $\partial E_{fit}/\partial V_t$ is inversely correlated with the amplitude of the phonon laser, which is experimentally controlled by the saturation ratio $r$ or the injection intensity $V_i$ \cite{SM}.

Besides, once $V_t$ increases to a level comparable to $V_i$, the measured $E_{fit}$ would deviate from the linear fitting and thus reduce the reconstruction accuracy. This deviation may arise from less stable oscillations caused by the competing locking of the input signal and the injection locking signal. Although increasing $V_i$ can broaden the measurement range of $V_t$ that helps promoting accuracy, it favors little for uncertainty reduction \cite{SM}. In this case, we have calculated the amplitude sensitivity to be 403.8 $\mu$V/($m\cdot{\mathrm{Hz}^{1/2}}$), with $\sigma_a$ = 9.5 mrad, which is comparable in order of magnitudes with the amplitude sensitivity of weak electric field signal we have detected in our former work \cite{Liu2021PhononLaser}.

Similarly, the fitted phase $\phi_{fit}$ also shows a favorable linear dependence on the given $\phi_t$, as plotted in Fig.~\ref{fig:3}(b), from which we acquire the phase sensitivity of 0.4(0.3) rad/$\sqrt{\mathrm{Hz}}$ with $\sigma_p$ = 98(65) mrad. We notice that the phase has a higher uncertainty than the other two components, because the long-term drift of $\omega_z$ mainly impacts the phase shift. This result can be optimized by improving the system stability.

\begin{figure}[t]
\centering
\includegraphics[width=0.45\textwidth]{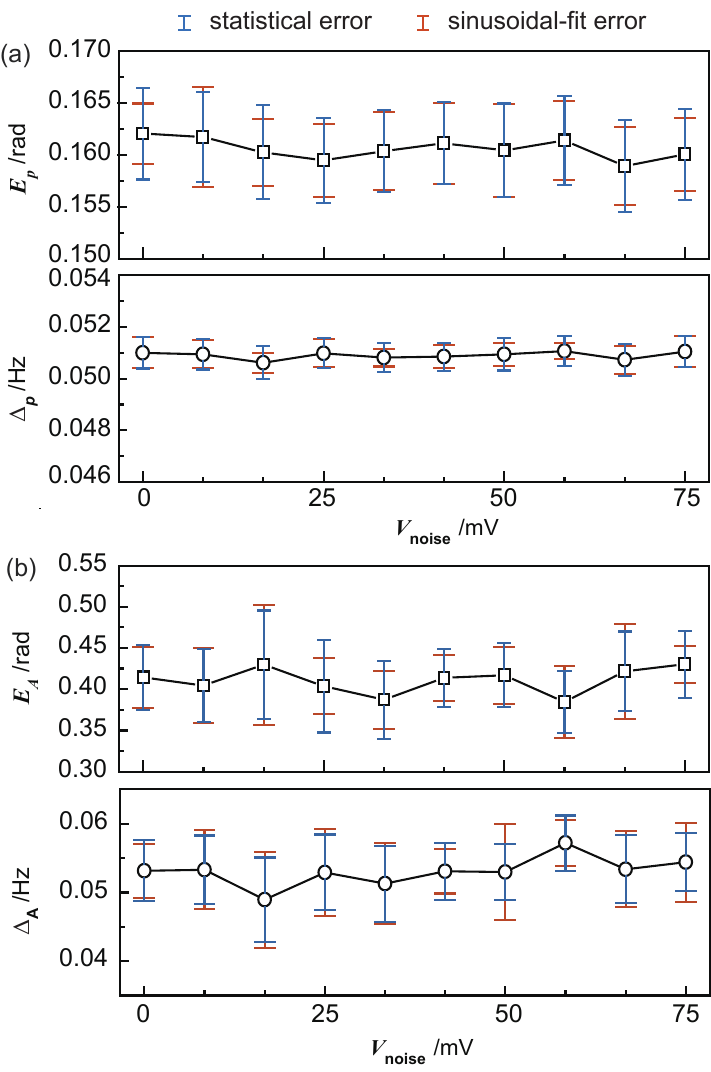}
\caption{The influence of different noise amplitudes on the fitted amplitude and frequency obtained via the evolution of the phonon laser's (a) phase and (b) amplitude with an input signal (1.25 mV, 0.05 Hz, 0), where the fitted amplitude/frequency obtained via fitting the evolution of phonon laser's phase and amplitude to Eq. (2) and Eq. (3) respectively are marked as $E_{p}$/$\Delta_{p}$ and $E_{A}$/$\Delta_{A}$.Blue error bars are the statistical standard error of 10 measurements, and red error bars are the sinusoidal fitting error.} \label{fig:4}
\end{figure}

The single-ion phonon laser beat frequency measurement system also demonstrates remarkable robustness against the applied Gaussian white noise. Fig.~\ref{fig:4} shows the evolution of the phonon laser's phase and amplitude with the noise increasing from 0 to 75 mV, which is much stronger than the input signal. Neither the fitted amplitudes, $E_{p}$, $E_{A}$, nor the fitted frequencies, $\Delta_{p}$, $\Delta_{A}$, obtained via respectively fitting the evolution of phonon laser's phase and amplitude to Eq. (2) and Eq. (3), show significant correlation with the noise intensity. Notably, both the statistical standard error (blue) and the sinusoidal fitting error (red) don't increase significantly with the noise amplitude, contrary to expectations. While the increased noise induces a measurable decrease in phonon laser amplitude, the measured phase stability remains exceptionally high \cite{SM}. Our approach is robust against noises since the noises predominantly affect the amplitude of a harmonic oscillator rather than the phase. This phase stability arises from injection locking to the phonon laser, which locks the oscillation frequency to the injection locking signal and thereby suppresses phase fluctuations. Crucially, parameters derived from sinusoidal fitting exhibit no significant correlation with the noise intensity, and the associated errors do not increase as expected. This unexpected noise immunity highlights the method's potential for high-precision measurements in complex, noisy environments. \\

We have experimentally demonstrated a practical and efficient scheme for measuring LFEF signals using an injection-locked single-ion phonon laser combined with the beat frequency technique, achieving a minimum detectable strength of LFEF of 61.5 $\mu$V/m, which is comparable in order of magnitudes to former trapped-ion electric-field sensing experiments \cite{InjectionLocking,Liu2021PhononLaser,Wei2022DCelectric}. Besides, different from our former work in electric field signal detection~\cite{PhysRevApplied.19.064062}, in the present work we directly apply the input LFEF signal to electrode without modulation, and realize the extraction of amplitude, frequency, and phase of the input signal in a single measurement.
The sensitivity of our scheme is determined by the measurement signal-to-noise ratio of the phonon laser, as well as the stability of the phonon laser's amplitude and phase. These quantities can be directly improved by enhancing the fluorescence collection efficiency and the long-term stability of the system, which could be realized through a more compact trap design and reduced secular frequency drift \cite{Johnson2016Active,Allcock2011Reduction}. In addition, enhancing the long-term stability of the system enables long-time continuous measurement, thereby further improving the frequency resolution.

Our approach enables the simultaneous extraction of the complete characteristics of the LFEF signal (i.e., amplitude, frequency, and phase) in a single measurement, making it more efficient than conventional counterparts. Moreover, our protocol operates near the secular frequency of the SET, which can be adjusted from 90 kHz to 300 kHz and can be expanded from a few kHz to the MHz regime in other types of ion traps~\cite{Romaszko2020, Xu2025}. Notably, this technique works without sideband cooling and relies entirely on sensitive optical detection. As a result, it is practical and easily applicable in trapped-ion systems for accurately detecting LFEF signals in noisy environments. Finally, the phase sensitivity of our scheme opens exciting possibilities for designing high-resolution mass spectrometry \cite{Desligniere2024Ultralong,Chen2024HighResolution} and detecting biochemical oscillations~\cite{Fei2018Design}.

Our study advances the development of LFEF sensors and inspires advancements in frequency modulation and conversion in nano-scale systems. Beyond trapped ions, the demonstrated transduction mechanism can be applied to other platforms, for example, the optomechanical microcavities \cite{Kuang2023Nonlinear,Zhang2022Dissipative}, electromechanical resonators \cite{Wen2020Coherent}, and optical tweezers \cite{Pettit2019Optical}, as well as to broader applications, such as long-wave communication \cite{Qu2024Review}, industrial and medical inspection \cite{Gajsek2016Review}, and meteorological monitoring \cite{Ahmadi2012Bioeffects}.\\



\noindent\textbf{Acknowledgements}\\
This work was supported by National Natural Science Foundation of China under Grant Nos. U25D9005, 12534020, 12304315, 12074346, 12074390, 12074280, 92265107, by Science and Technology Projects in Guangzhou under Grant Nos. 202201011727 and 2023A04J0050, by Guangdong Provincial Quantum Science Strategic Initiative under Grand No. GDZX2305004 and No. GDZX2505001, by Nansha Senior Leading Talent Team Technology Project under Grant No. 2021CXTD02, by Science and Technology Project of SGCC 52120025005C-338-WLCY,
and by the Special Project for Research and Development in Key Areas of Guangdong Province under Grant No. 2020B0303300001.\\

\bibliography{biblio}



\end{document}